\documentclass[fleqn,usenatbib]{mnras}

\usepackage[english]{babel}
\usepackage[usenames]{color}

\usepackage{newtxtext,newtxmath}
\usepackage[T1]{fontenc}
\usepackage{graphicx}	
\usepackage{amsmath}	

\title[Physical and chemical vertical structure of accretion disks]{Physical and Chemical Vertical Structure of Magnetostatic  Accretion Disks of Young Stars}

\author[S. A. Khaibrakhmanov, et al.]{
Sergey A. Khaibrakhmanov$^{1,2}$\thanks{E-mail: khaibrakhmanov@csu.ru},
Alexander E. Dudorov$^{2,1}$\thanks{deceased},
Anton I. Vasyunin$^{1}$,
Mikhail Yu. Kiskin$^{1}$
\\
$^{1}$Ural Federal University, 51 Lenin st., Ekaterinburg 620000, Russia\\
$^{2}$Chelyabinsk State University, 129 Br. Kashirinykh st., Chelyabinsk 454001, Russia
}

\date{Accepted 28.08.2021}
\pubyear{2021}

\begin{document}

\label{firstpage}
\pagerange{\pageref{firstpage}--\pageref{lastpage}}
\maketitle

\begin{abstract}
The vertical structure of the accretion disks of young stars with fossil large-scale magnetic field
is studied. The equations of magnetostatic equilibrium of the disk are solved taking into account the
stellar gravity, gas and magnetic pressure, turbulent heating, and heating by stellar radiation. The modelled physical structure of the disk is used to simulate its chemical structure, in particular, to study the
spatial distribution of CN molecules. The disk of the typical T Tauri star is considered. Simulations show that the temperature within the disk in the region $r<50$~au decreases with height and density profiles are steeper than in the isothermal case. Outside the `dead' zone, vertical profiles of the
azimuthal component of the magnetic field are nonmonotonic, and the magnetic field strength maximum is reached within the disk. The magnetic pressure gradient can cause an increase in the disk
thickness in comparison with the hydrostatic one. 
The CN molecule concentration is maximum near the photosphere and in the disk atmosphere where the magnetic field strength at chosen parameters is $\sim 0.01$~G. Measurements of the Zeeman splitting of CN lines in the submm range can be used to determine the magnetic field strength in these regions of accretion disks.
\end{abstract}

\begin{keywords}
accretion disks -- magnetic field -- magnetohydrodynamics (MHD) -- astrochemistry
\end{keywords}

\section*{Introduction}
Accretion disks of young stars (ADYS) are geometrically thin optically thick gas--dust disks with characteristic
sizes of $100-1000$~au and masses $0.001-0.1\,M_{\odot}$. 
During the evolution, ADYS become protoplanetary disks with conditions favorable for planet formation. 

Polarization mapping of thermal radiation of ADYS and observations of outflows and jets suggest that
a large-scale magnetic field exists in disks~\cite[see review by][]{cmf}. The spatial resolution and sensitivity of available instruments do not allow detailed conclusions on themagnetic field geometry in ADYS. The magnetic field strength in ADYS can be indirectly estimated by the remanent magnetization of meteorites of the
solar system; its direct measurements using the Zeeman effect are still challenging. A promising direction
is the measurement of the Zeeman splitting of CN molecule lines in the submm range~\citep{vlemmings19}.

An analysis of observations of star formation regions and numerical simulations of star formation
resulting from the collapse of magnetic rotating cores of molecular clouds show that the magnetic
field in ADYS is of fossil nature~\citep[see][]{fmft}. Within the theory of the fossil magnetic field,~\cite{DKh14} developed a magnetohydrodynamic (MHD) model of ADYS. Using it, it was in particular
shown that the magnetic field can be dynamically strong in some disk regions.

In the present paper, the approach by Dudorov and Khaibrakhmanov is developed, and the effect of
the magnetic field on the vertical structure of the ADYS is considered. Chemical modeling of the ADYS is
performed and the spatial distribution of CN molecules in the disk is determined.

\section{Problem statement and main equations}
\label{sec:problem}
Let us consider a low-mass steady-state geometrically thin and optically thick accretion disk with a fossil
large-scale magnetic field. The inner radius of the disk is defined by the stellar magnetosphere radius, the outer radius of the disk is defined as the contact boundary with the interstellar medium. The disk surface
represents the boundary of its photosphere which is also a contact one. 

To study the dynamics of the ADYS with large-scale magnetic field, we use MHD equations taking into
account the stellar gravity, turbulent viscosity, radiative thermal conductivity, and magnetic field diffusion.
In the approximation of the steady-state geometrically thin disk, this system of equations is split into
two independent subsystems describing the radial and vertical disk structure. 

The basic system of equations for simulating the ADYS radial structure with magnetic field was
derived by Dudorov and Khaibrakhmanov and was described in detail in~\cite{Kh17}. Following~\cite{SS73}, it is supposed that the main mechanism of angular momentum transport in the differential
rotating disk is turbulence, and the main gas heating mechanism is turbulent friction. In addition to Shakura
and Sunyaev equations, equations of collisional and thermal ionization are solved, as well as the induction
equation taking into account Ohmic dissipation and magnetic ambipolar diffusion, buoyancy, and the
Hall effect. 

The equations of the magnetostatic equilibrium of the ADYS can be written in cylindrical coordinates $(r,\,0,\,z)$~\citep{KhD21}:
\begin{eqnarray}
\frac{\partial p}{\partial z} &=& \rho g_z + \frac{\partial}{\partial z}\left(\frac{B_{\varphi}^2}{8\pi}\right),\label{eq:magnetostatic}\\
\kappa_{\rm r}\frac{\partial T}{\partial z} &=&  \mathcal{F}_z,\label{eq:dT_dz}\\
\frac{\partial \mathcal{F}_z}{\partial z} &=& \Gamma_{\rm turb},\label{eq:dQturb_dz}\\
\frac{\partial^2 B_\varphi}{\partial z^2} &=& -\frac{3}{2}\frac{v_{\rm k}B_z}{\eta}\frac{z}{r^2} \label{eq:induction_phi},
\end{eqnarray}
where $g_z$~is the vertical component of the stellar gravity, $\kappa_{\rm r}$~is the radiative thermal conductivity, $\mathcal{F}_z$~is the flux density of stellar radiation in the vertical direction, $\Gamma_{\rm turb} = -\alpha p rd\Omega/dr$~is the gas volume heating rate due to turbulent friction, $\alpha$~is the Shakura and Sunyaev turbulent parameter, $\Omega$~is the gas angular velocity, $v_{\rm k}$~is the Keplerian velocity, $\eta$~is the magnetic diffusivity, $B_\varphi$ and $B_z$~are the azimuthal and vertical components of the magnetic field.

Equations (\ref{eq:magnetostatic}--\ref{eq:induction_phi}) imply that turbulent friction is a single heating source within the disk. An additional heating mechanism of upper disk layers is the absorption of stellar rediation. The stellar radiation flux is incident on the disk surface at a very small angle; therefore, for simplicity, it can be considered that it is completely absorbed in the optically thin disk atmosphere. The atmosphere temperature $T_{\rm a}$ can be determined from the heat balance condition:
\begin{equation}
\sigma T_{\rm a}^4 = f\frac{L_\star}{4\pi r^2},
\end{equation}
where $L_\star$~is the stellar luminosity and $f<1$~is the geometrical factor defining the fraction of radiation
absorbed by the disk at a given distance. The value of $f$ depends on the disk surface shape and, generally
speaking, is a priori unknown. For example, according to detailed simulations by~\cite{akimkin12}, $f$ can vary from $0.015$ to $0.15$. In the present study, a constant characteristic value of $0.05$ is taken. 

To solve Eqs.~(\ref{eq:magnetostatic}--\ref{eq:induction_phi}), five boundary conditions should be set. Let us consider the region from the equatorial plane of the disk, $z=0$, to its photosphere boundary, $z_{\rm{s}}$, characterized by the optical thickness $\tau=2/3$. 
Due to the equatorial plane symmetry $\mathcal{F}_z=0$, $B_\varphi=0$. At the disk surface $T=T_{\rm eff}$, $\mathcal{F}_z = \sigma T_{\rm eff}^4$, where $T_{\rm eff}$~is the effective disk temperature. The gas pressure $p_{\rm s}$ over disk corresponds to the pressure of the central region of the molecular cloud core with a characteristic density of $10^9$~cm$^{-3}$ and temperature of $20$~K. The magnetic field strength at the surface, $B_{\rm s}$, is defined in terms of the plasma parameter $\beta=8\pi p_{\rm s}/B_{\rm s}^2$.

\section{Method for solving equations and model parameters}
\label{sec:methods}
The induction Eq.~(\ref{eq:induction_phi}) can be solved analytically. For the chosen boundary conditions, we derive~\citep{KhD21}
\begin{equation}
 B_{\varphi}(r,\,z) = B_{\rm s}\frac{z}{z_{\rm s}} + \frac{1}{4}\frac{v_k z}{\eta}B_z\left[\left(\frac{z}{r}\right)^2 - \left(\frac{z_{\rm s}}{r}\right)^2\right].\label{eq:Bphi_I}
\end{equation}

The system of the first-order ordinary differential equations (\ref{eq:magnetostatic}--\ref{eq:dQturb_dz}) 
is solved numerically by the fourth-order Runge--Kutta method with automatic step selection for a relative accuracy of $10^{-4}$.  

In this paper, the structure of the accretion disk of a T Tauri star of solar mass with an accretion
rate of $10^{-8}\,M_{\odot}/\mathrm{yr}$ and turbulence parameter $\alpha=0.01$ is simulated. The characteristic radial distances from star $r=0.25$, $1$, $10$ and $50$~au are considered. The range $0.25 < r < 50$~au corresponds to the `dead' zone of the disk, where the magnetic field diffusion prevents its generation. The equation coefficients at each $r$ were set from the solution of equations of the radial disk structure with the help of Dudorov and Khaibrakhmanov model. The induction equation is solved taking into account Ohmic dissipation, $\eta = \nu_{\rm m} = c^2 / (4\pi\sigma_{\rm e})$,  
where $\sigma_{\rm e}(x)$~is the plasma Coulomb conductivity, and $x$~is the ionization fraction. 
Ohmic dissipation is the main dissipative MHD effect in `dead' zones of accretion disks~\citep[see][]{Kh17}. The degree of ionization was calculated taking into account radiative recombinations and recombinations on dust particles with a characteristic average radius of~$0.1\,\mu$m, for standard rates of ionization by cosmic rays, X-rays, and radioactive elements. At the chosen parameters, the magnetic Reynolds number, $Re_{\rm m}=v_{\rm k}H/\nu_m$, characterizing the efficiency of magnetic field diffusion, is 3, 68 and 8425 at chosen distances of 0.25, 10 and 50 au, respectively. The plasma beta on the surface is taken equal to unity.

The results of calculations of the physical structure of the disk were used for simulating its chemical
composition using the MONACO numerical code~\citep{monaco13, monaco17}. Chemical kinetic equations describing 6002 chemical reactions between 664 atoms and molecules consider gas-phase reactions, reactions on the surface
and in the bulk of icy mantles of dusty particles. It is assumed that the dust has a standard chemical
composition and is well mixed with gas. The coefficients of the interaction of atoms and molecules with
dust particles are taken from~\cite{hasegawa}. The initial chemical composition was set as a result of simulations of the chemical evolution of the typical molecular cloud for $10^6$ years. The disk chemical evolution was calculated for a time interval of $10^6$ years. The system of chemical balance equations was solved using the DVODE integrator implementing the implicit Adams--Gear method.

\section{Results}
\label{sec:results}

\begin{figure*}
   \centering
  \includegraphics[trim = 0 3cm 0 0, width=0.79\textwidth]{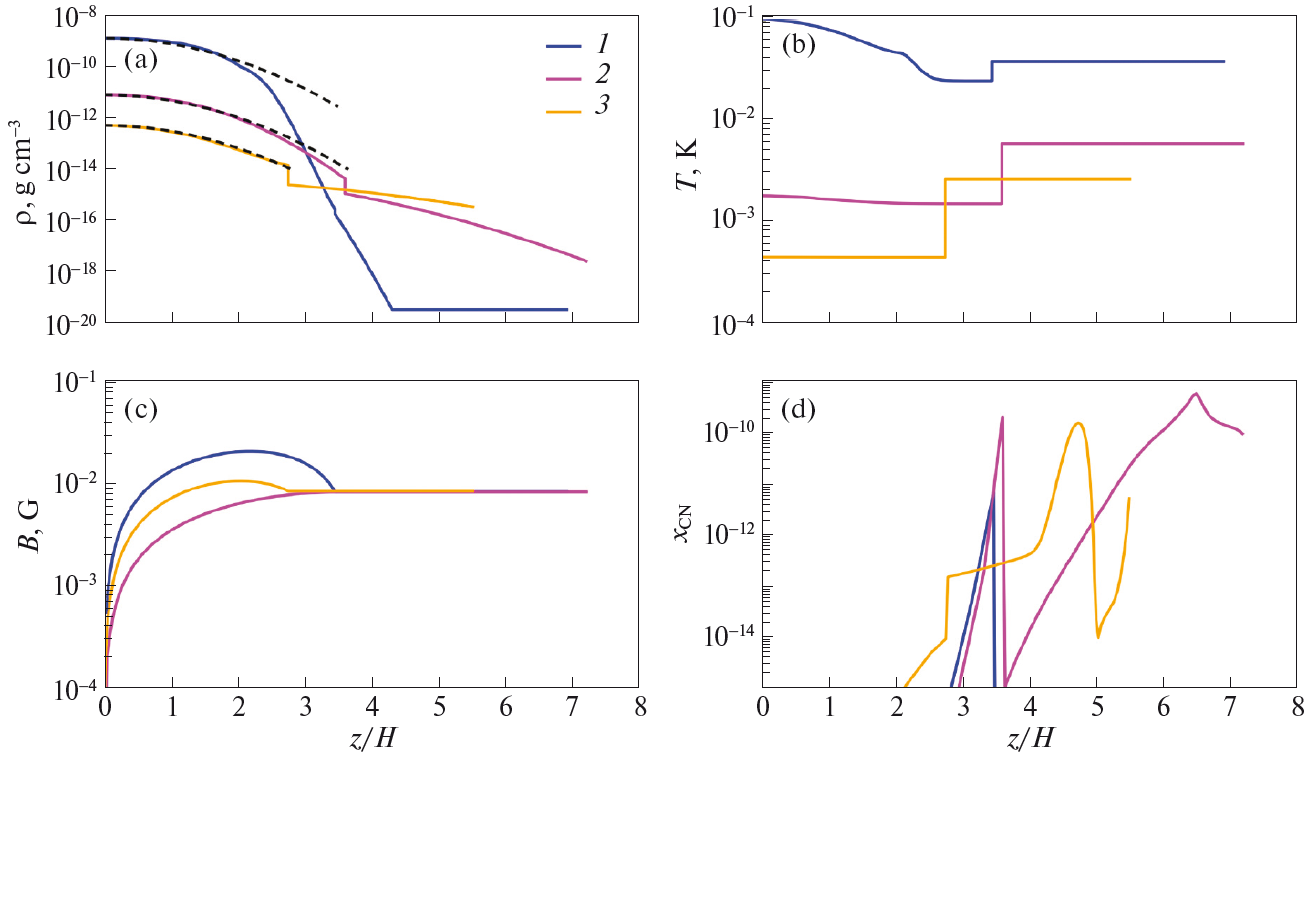}
   \caption{Vertical profiles of density ({\it a}), temperature ({\it b}), magnetic field strength ({\it c}), and abundance of CN molecules in the gas phase relative to hydrogen ({\it d}) at various radial distances from the star. Curves `1', `2', and `3' correspond to $r=0.25$, $10$, and $r=50$~au, respectively. Dashed curves in ({\it a}) are the corresponding profiles for the case of the disk with a height-constant temperature, $T=T(z=0)$.}
   \label{fig:profiles}
\end{figure*}

Figure~\ref{fig:profiles} shows the vertical profiles of the density, temperature, magnetic field strength, and CN molecule concentration in the gas phase relative to hydrogen. According to the performed calculations,
at $r=0.25$, $10$ and $50$~au the photosphere is at heights $z_{\rm s}=3.8$, $2.7$ and $2.8\,H$ respectively. Here $H$~is the scale height determined by the temperature in the equatorial plane.

Figure~\ref{fig:profiles}a shows that the density profiles in the region $r<50$ au, $2H < z <z_{\rm s}$ (curves `1' and `2') are steeper than in the isothermal case. This is due to the fact that the temperature inside the disk in this range decreases with height (see. Fig.~\ref{fig:profiles}b). The largest difference is observed in the innermost region, $r=0.25$~au, where the temperature decreases from $\approx 1000$~K in the equatorial plane to $\approx 300$~K on the surface, and the density near the photosphere, $\rho\approx 10^{-16}\,\mathrm{g}\,\mathrm{cm}^{-3}$, is lower than the hydrostatic one by 4--5 orders of magnitude. In
the outer disk region, $r=50$~au (curve `3'), the temperature is constant in height and the density profile is
identical to the isothermal one. On the contact disk surface, $z=z_{\rm s}$ the temperature abruptly increases
when going from the disk whose heating is caused by turbulent friction to its atmosphere whose heating is
controlled by stellar radiation. In the disk atmosphere, $z>z_{\rm s}$, the temperature is constant and the density exponentially decreases with height down to the interstellar medium density.

Figure~\ref{fig:profiles}c shows that $B_{\varphi}$ increases with height $z$ outside the dead zone, at $r=0.25$ and $50$~au, near the equatorial plane and decreases near the disk surface, i.e., $B_{\varphi}$ takes a maximum value within the disk, e.g., $B_{\varphi}\approx 0.01$~G at $z\approx 2\,H$, $r=50$~au. In this case, the magnetic pressure gradient near the disk surface results in an increase in the characteristic disk thickness in comparison with the hydrostatic one~\citep{KhD21}.

According to Fig.~\ref{fig:profiles}d, CN molecules are distributed along $z$ highly inhomogeneously. The CN abundance is maximum, $x_{\rm CN}\sim (10^{-10}-10^{-9})$, near the disk surface, $z\sim z_{\rm s}\approx 3\,H$, as well as in its atmosphere. Maximum values of $x_{\rm CN}$ in these disk regions are close to the results obtained in~\cite{cazzoletti18} for accretion disks of young T~Tauri stars.

\section{Conclusions}
\label{sec:conclude}
Physical and chemical vertical structure of the ADYS with fossil large-scale magnetic was simulated. The
model of the vertical disk structure, developed in~\cite{KhD21}, was complemented for considering the effect of disk atmosphere heating by stellar radiation.

The simulations show that in the inner region, $r<50$~au, the disk has a smaller characteristic thickness,
than in the isothermal case. The magnetic field profiles outside the `dead' zone are nonmonotonic, the
maximum strength $B_\varphi$ is reached inside the disk. In this case, the magnetic pressure gradient can lead to an increase in the effective disk thickness in comparison with the hydrostatic one~\citep{KhD21}. Other studies, as a rule, considered the case where the magnetic field strength monotonically increases to the surface, and the magnetic pressure gradient `compresses' the disk~\citep[see][]{lizano17}. Generally speaking, both cases are possible depending on surface conditions.

CN molecule abundance in the gas phase is maximum $x_{\rm CN}\sim (10^{-10}-10^{-9})$, in the disk photosphere and atmosphere. Measurements of the Zeeman splitting of CN lines can be used to determine the magnetic field
strength in these regions, where the magnetic field strength for chosen parameters is $\sim 0.01$~G. Quantitative interpretation of available and future observations requires more detailed two-dimensional MHD simulation of the disk.

{\bf Acknowledgments.}
S.~A. Khaibrakhmanov acknowledges the support of the Government of the Russian Federation and the Ministry of
Higher Education and Science of the Russian Federation project no. 075-15-2020-780 (N13.1902.21.0039, contract no. 780--10). A.~I.~Vasyunin acknowledges the support of the Russian Science Foundation (project no. 19-72-10012). The authors thank anonymous referee for some useful comments.

\bibliographystyle{mnras}
\bibliography{bib}

\begin{thebibliography}{}
\makeatletter
\relax
\def\mn@urlcharsother{\let\do\@makeother \do\$\do\&\do\#\do\^\do\_\do\%\do\~}
\def\mn@doi{\begingroup\mn@urlcharsother \@ifnextchar [ {\mn@doi@}
  {\mn@doi@[]}}
\def\mn@doi@[#1]#2{\def\@tempa{#1}\ifx\@tempa\@empty \href
  {http://dx.doi.org/#2} {doi:#2}\else \href {http://dx.doi.org/#2} {#1}\fi
  \endgroup}
\def\mn@eprint#1#2{\mn@eprint@#1:#2::\@nil}
\def\mn@eprint@arXiv#1{\href {http://arxiv.org/abs/#1} {{\tt arXiv:#1}}}
\def\mn@eprint@dblp#1{\href {http://dblp.uni-trier.de/rec/bibtex/#1.xml}
  {dblp:#1}}
\def\mn@eprint@#1:#2:#3:#4\@nil{\def\@tempa {#1}\def\@tempb {#2}\def\@tempc
  {#3}\ifx \@tempc \@empty \let \@tempc \@tempb \let \@tempb \@tempa \fi \ifx
  \@tempb \@empty \def\@tempb {arXiv}\fi \@ifundefined
  {mn@eprint@\@tempb}{\@tempb:\@tempc}{\expandafter \expandafter \csname
  mn@eprint@\@tempb\endcsname \expandafter{\@tempc}}}

\bibitem[\protect\citeauthoryear{{Akimkin}, {Pavlyuchenkov}, {Launhardt}  \&
  {Bourke}}{{Akimkin} et~al.}{2012}]{akimkin12}
{Akimkin} V.~V.,  {Pavlyuchenkov} Y.~N.,  {Launhardt} R.,   {Bourke} T.,  2012,
  \mn@doi [Astronomy Reports] {10.1134/S1063772912120013}, \href
  {https://ui.adsabs.harvard.edu/abs/2012ARep...56..915A} {56, 915}

\bibitem[\protect\citeauthoryear{{Cazzoletti}, {van Dishoeck}, {Visser},
  {Facchini}  \& {Bruderer}}{{Cazzoletti} et~al.}{2018}]{cazzoletti18}
{Cazzoletti} P.,  {van Dishoeck} E.~F.,  {Visser} R.,  {Facchini} S.,
  {Bruderer} S.,  2018, \mn@doi [\aap] {10.1051/0004-6361/201731457}, \href
  {https://ui.adsabs.harvard.edu/abs/2018A&A...609A..93C} {609, A93}

\bibitem[\protect\citeauthoryear{{Dudorov} \& {Khaibrakhmanov}}{{Dudorov} \&
  {Khaibrakhmanov}}{2014}]{DKh14}
{Dudorov} A.~E.,  {Khaibrakhmanov} S.~A.,  2014, \mn@doi [\apss]
  {10.1007/s10509-014-1900-4}, \href
  {https://ui.adsabs.harvard.edu/abs/2014Ap&SS.352..103D} {352, 103}

\bibitem[\protect\citeauthoryear{{Dudorov} \& {Khaibrakhmanov}}{{Dudorov} \&
  {Khaibrakhmanov}}{2015}]{fmft}
{Dudorov} A.~E.,  {Khaibrakhmanov} S.~A.,  2015, \mn@doi [Advances in Space
  Research] {10.1016/j.asr.2014.05.034}, \href
  {https://ui.adsabs.harvard.edu/abs/2015AdSpR..55..843D} {55, 843}

\bibitem[\protect\citeauthoryear{{Hasegawa} \& {Herbst}}{{Hasegawa} \&
  {Herbst}}{1993}]{hasegawa}
{Hasegawa} T.~I.,  {Herbst} E.,  1993, \mn@doi [\mnras]
  {10.1093/mnras/261.1.83}, \href
  {https://ui.adsabs.harvard.edu/abs/1993MNRAS.261...83H} {261, 83}

\bibitem[\protect\citeauthoryear{{Khaibrakhmanov} \&
  {Dudorov}}{{Khaibrakhmanov} \& {Dudorov}}{2019}]{cmf}
{Khaibrakhmanov} S.~A.,  {Dudorov} A.~E.,  2019, in Trudy 48 mezhdunarodnoi
  konferentsii `Fizika Kosmosa' (Proc. 48th International Student Scientific
  Conference `Physics of Space'). Yekaterinburg, Russia, pp 92--111

\bibitem[\protect\citeauthoryear{{Khaibrakhmanov} \&
  {Dudorov}}{{Khaibrakhmanov} \& {Dudorov}}{2021}]{KhD21}
{Khaibrakhmanov} S.~A.,  {Dudorov} A.~E.,  2021, \mn@doi [Chelyabinsk Physical
  and Mathematical Journal] {10.47475/2500-0101-2021-16105}, 6, 53

\bibitem[\protect\citeauthoryear{{Khaibrakhmanov}, {Dudorov}, {Parfenov}  \&
  {Sobolev}}{{Khaibrakhmanov} et~al.}{2017}]{Kh17}
{Khaibrakhmanov} S.~A.,  {Dudorov} A.~E.,  {Parfenov} S.~Y.,   {Sobolev} A.~M.,
   2017, \mn@doi [\mnras] {10.1093/mnras/stw2349}, \href
  {https://ui.adsabs.harvard.edu/abs/2017MNRAS.464..586K} {464, 586}

\bibitem[\protect\citeauthoryear{{Lizano}, {Tapia}, {Boehler}  \&
  {D'Alessio}}{{Lizano} et~al.}{2016}]{lizano17}
{Lizano} S.,  {Tapia} C.,  {Boehler} Y.,   {D'Alessio} P.,  2016, \mn@doi
  [\apj] {10.3847/0004-637X/817/1/35}, \href
  {https://ui.adsabs.harvard.edu/abs/2016ApJ...817...35L} {817, 35}

\bibitem[\protect\citeauthoryear{{Shakura} \& {Sunyaev}}{{Shakura} \&
  {Sunyaev}}{1973}]{SS73}
{Shakura} N.~I.,  {Sunyaev} R.~A.,  1973, \aap, \href
  {https://ui.adsabs.harvard.edu/abs/1973A&A....24..337S} {24, 337}

\bibitem[\protect\citeauthoryear{{Vasyunin} \& {Herbst}}{{Vasyunin} \&
  {Herbst}}{2013}]{monaco13}
{Vasyunin} A.~I.,  {Herbst} E.,  2013, \mn@doi [\apj]
  {10.1088/0004-637X/762/2/86}, \href
  {https://ui.adsabs.harvard.edu/abs/2013ApJ...762...86V} {762, 86}

\bibitem[\protect\citeauthoryear{{Vasyunin}, {Caselli}, {Dulieu}  \&
  {Jim{\'e}nez-Serra}}{{Vasyunin} et~al.}{2017}]{monaco17}
{Vasyunin} A.~I.,  {Caselli} P.,  {Dulieu} F.,   {Jim{\'e}nez-Serra} I.,  2017,
  \mn@doi [\apj] {10.3847/1538-4357/aa72ec}, \href
  {https://ui.adsabs.harvard.edu/abs/2017ApJ...842...33V} {842, 33}

\bibitem[\protect\citeauthoryear{{Vlemmings} et~al.,}{{Vlemmings}
  et~al.}{2019}]{vlemmings19}
{Vlemmings} W.~H.~T.,  et~al., 2019, \mn@doi [\aap]
  {10.1051/0004-6361/201935459}, \href
  {https://ui.adsabs.harvard.edu/abs/2019A&A...624L...7V} {624, L7}

\makeatother
\end{thebibliography}

\end{document}